# Distributed laser energy deposition enables fog-resilient optical communications

Malte C. Schroeder[1,2], Victor Moreno[1], Christophe Coreixas[3,4], Jonas Latt[4] and Jean-Pierre Wolf[1]

(1) *Applied Physics Department, University of Geneva, 1211 Geneva 4 (Switzerland)*

*(2) Photonics for Ultrafast Laser Science (PULS), Ruhr-University-Bochum, 44801 Bochum (Germany)*

(3) *Institute for Advanced Study, Beijing Normal-Hong Kong Baptist University, 519087 Zhuhai (China)*

(4) *Computer Science Department, University Geneva, 1211 Geneva 4 (Switzerland)*

## Abstract

Fog and clouds remain major obstacles to free-space optical (FSO) communications because micron-sized droplets can impose tens of decibels of attenuation. Ultrashort laser filamentation can transiently clear fog in laboratory-scale experiments, but its extension to atmospheric-scale propagation has remained uncertain. Here we demonstrate laser-assisted fog clearing under atmospheric-scale propagation conditions over a 140-m optical path using a high-energy, high-average-power picosecond laser and weak focusing to sustain extended multifilamentation. A steady-state transparent channel approximately 2 cm in diameter restores up to 12 dB of transmission for an independent optical beam. Increasing the clearing-laser pulse energy primarily extends the multifilament region while maintaining a nearly constant longitudinally averaged deposited energy per unit length of ~2 mJ $m^{-1}$ over the measured bundle length. This low average deposition, together with self-sustained filament propagation through dense fog, provides a favorable scaling route toward weather-resilient optical links.

Free-space optical (FSO) communications offer high bandwidth, low latency and intrinsic resistance to eavesdropping for terrestrial, airborne and satellite networks, but remain strongly constrained by adverse weather (Muhammad et al., 2007; Chen et al., 2021; Jahid et al., 2022). Fog and clouds, composed of micron-sized water droplets, can impose attenuations of tens of decibels and interrupt links over distances from hundreds of meters to several kilometers. Across Europe, average cloud coverage ranges from approximately 55% to 60% of the time, reaching up to 80% in Northern Europe (Karlsson and Devasthale, 2018). Hybrid RF/FSO systems, site diversity and adaptive routing can improve availability, while structured-light and obstruction-free-channel approaches can improve transmission through scattering or obstructed paths (Ali Reza et al., 2023; Wang et al., 2023), but do not remove the atmospheric obstruction itself.

Ultrashort laser filamentation offers a different approach by actively creating transparent channels through droplet clouds. Previous laboratory experiments demonstrated laser-assisted optical transmission through sub-meter-scale fog volumes (De La Cruz et al., 2016; Schimmel et al., 2018; Schroeder et al., 2020; Yan et al., 2020; Goffin et al., 2022). At high repetition rates, clearing is expected to involve an initial regime dominated by direct laser-droplet interactions, followed by a steady-state regime sustained by repeated shock waves and cumulative air heating that inhibit droplet refilling (Lahav et al., 2014; Schubert et al., 2016; Walch et al., 2021; Schroeder et al., 2022).

Scaling this concept to atmospheric propagation is not simply a matter of increasing pulse energy. Efficient clearing requires energy deposition throughout the obstructing layer, whereas a conventionally focused beam loses its ballistic component through scattering and deposits most of its energy near focus. Filamentation can instead transport high optical intensities while enabling distributed energy deposition over extended propagation distances (Kasparian et al., 2003; Rodriguez et al., 2004; Chin et al., 2005; Bergé et al., 2007; Couairon and Mysyrowicz, 2007; Durand et al., 2013; Point et al., 2014; Houard et al., 2016; Loescher et al., 2023; Walch et al., 2023).

Here we address this scaling question experimentally. A kilohertz picosecond laser, expanded to a 40-cm aperture and weakly focused, was propagated over a 140-m path through dense fog generated at 50 m and 100 m standoff, and the channel it opened was probed by an independent co-propagating optical beam. Three quantities are reported: the transmission of the clearing laser through the fog, the establishment and long-time maintenance of the transparent channel, and the energy deposited per unit propagation length. Together, they identify the average deposition per unit length, rather than the total pulse energy, as the quantity that governs how far a cleared channel can be extended.

## Self-sustained propagation through dense fog

To clear fog over atmospheric distances, the clearing laser must itself remain capable of propagating through the droplet cloud. The characterization of the nonlinear propagation of this laser system over long distances has already been presented in detail in Walch et al. (2023b). Figure 1 shows the transmission of the filamenting laser as a function of pulse energy for different fog attenuations and propagation geometries. Transmission values are referenced to clear-air propagation and therefore include both nonlinear propagation losses and scattering by the fog droplets.

For all investigated configurations, the transmission increases rapidly with pulse energy before approaching the clear-air value. Pulse energies of approximately 100 mJ are sufficient to recover nearly the clear-air transmission of the clearing laser, even for the highest fog attenuation investigated (13.5 dB).

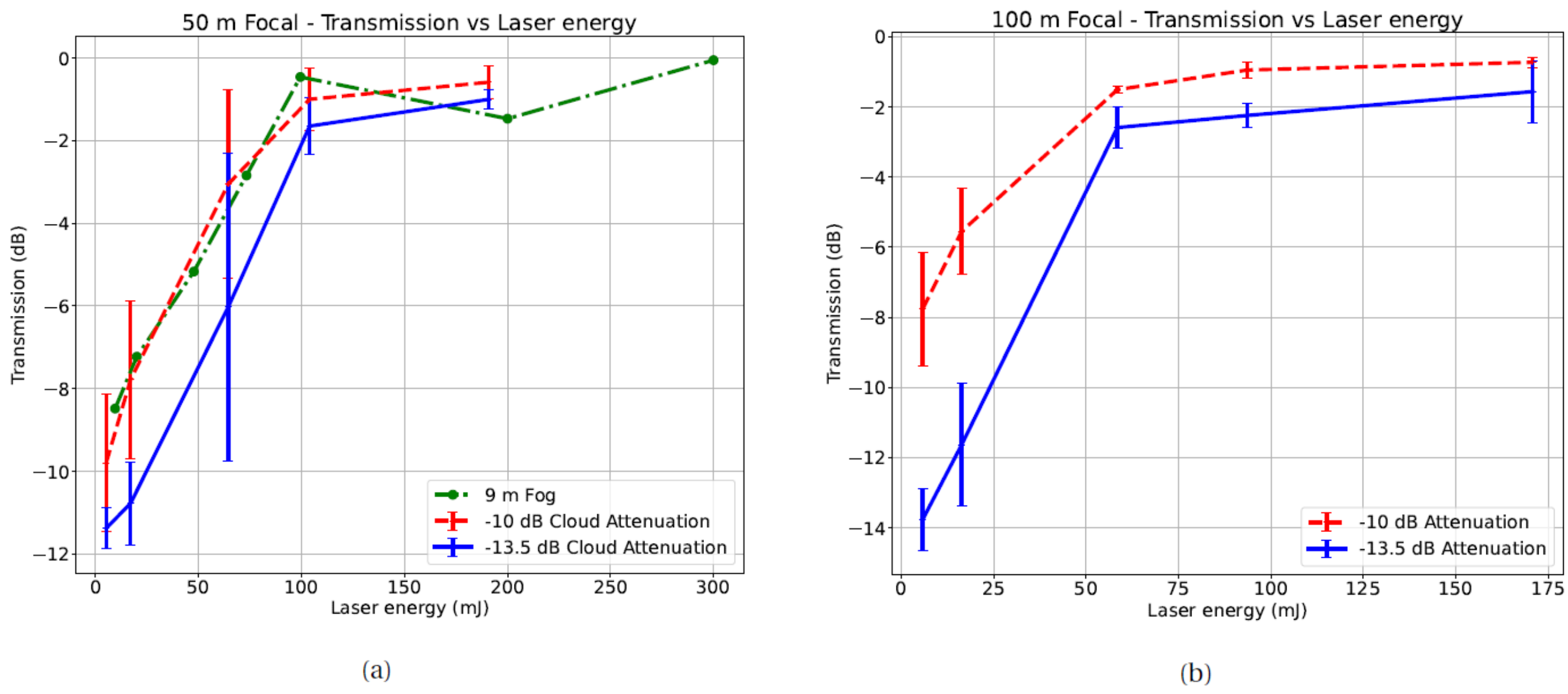

**Figure 1**: Transmission of the high-power laser through fog, for different configurations and droplet densities (expressed in dB). (a) 3 m- and 9 m- long cloud chamber at 50 m, (b) 3-m long cloud chamber at 100 m. The cloud chamber length is 3 m, unless specified in the figure

Notice that an attenuation of 13.5 dB corresponds to a droplet concentration as high as ~ $2.7 \times 10^4$ $cm^{-3}$ (LWC ~ 1.7 g $m^{-3}$) (calculated using Mie theory for the mean droplet size). In comparison, a standard natural fog contains some tens to some hundreds droplets per $cm^3$ (García-García et al., 2002).

Beyond this energy, the transmission reaches a plateau, with the residual loss remaining below approximately 1-2 dB once steady-state clearing has been established. Similar behavior is observed for both the 50 m and 100 m focusing geometries, as well as for the 3 m and 9 m fog chambers, demonstrating the robustness of the clearing process over a broad range of experimental conditions. However, producing a homogeneous and stable fog distribution over the 9 m length using multiple nebulizers proved experimentally challenging, and all subsequent investigations were therefore performed with the 3 m fog configuration, which provided superior stability and reproducibility.

These results demonstrate that, once the transparent channel has been established, the filamenting beam becomes self-sustained and continues to deliver energy throughout the fog layer instead of being progressively disrupted by multiple scattering (Courvoisier et al., 2003b; Méjean et al., 2005; Frigerio et al., 2024). This self-sustained propagation is a prerequisite for maintaining the transparent channel. The mechanisms governing its establishment and long-term maintenance are examined in the following section.

## Formation of the transparent channel

To investigate the temporal establishment of the transparent channel, a low-power He-Ne laser was spatially superimposed with the filamenting beam and served as a surrogate communication beam throughout the experiments. Since the He-Ne beam does not modify the fog, its transmission directly monitors the optical transparency created by the clearing laser.

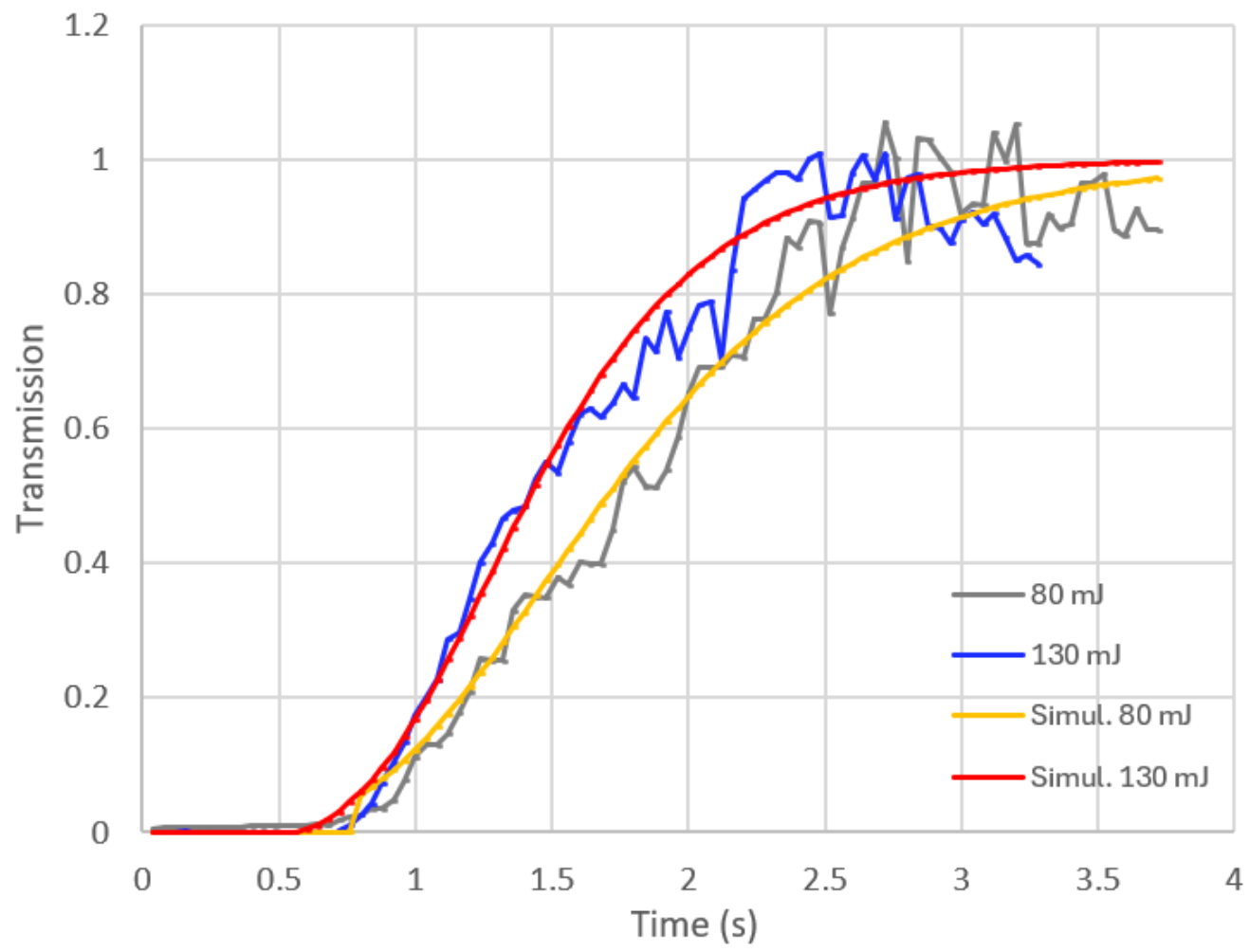

**Figure 2**: Establishment dynamics of the transparent channel. Grey and blue curves are experimental results for laser energies of 80 mJ and 130 mJ, respectively, while yellow and red curves show the results of the numerical simulations for the same energies

Figure 2 shows that the transparent channel is established in two successive stages. Immediately after the clearing laser is switched on, the He-Ne transmission increases progressively during a transient regime lasting a few seconds, depending on the laser pulse energy and the initial fog attenuation. For a dense fog of 12 dB attenuation in the 3 m chamber located at 100 m, the initial clearing is completed in about 1.5 s (1500 shots) at 130 mJ and 2.4 s (2400 shots) at 80 mJ. This first stage corresponds to the progressive formation of the transparent channel through laser-induced droplet shattering and evaporation (Courvoisier et al., 2003a; Lindinger et al., 2004; Jeon et al., 2015; Zhang et al., 2019; Peña et al., 2021; Goffin et al., 2022). Once the channel has been established, the transmission reaches a stable plateau, indicating the onset of a steady-state clearing regime.

To interpret this initial channel formation, numerical simulations considering only laser-induced droplet explosion/evaporation (see Methods) were compared with the experimental measurements. The simulations reproduce the measured transmission dynamics with good agreement, despite neglecting droplet replenishment into the cleared channel. They further indicate that the characteristic opening time is primarily governed by the number of filaments sustained within the bundle --approximately 10 filaments at 130 mJ and 5 filaments at 80 mJ-- rather than by the total laser pulse energy itself.

In a second stage, repeated shock waves and cumulative air heating generated by the 1-kHz filamenting beam (Lahav et al., 2014; Schubert et al., 2016; Walch et al., 2021; Schroeder et al., 2022) continuously prevent droplets from refilling the channel, allowing a transparent channel to be continuously maintained throughout laser operation. Successive pulses deepen the long-lived density depression generated in the wake of the filament, thereby enhancing the amplitude of the resulting shock wave and its ability to expel droplets from the beam path (Walch et al., 2021). For the smallest droplets in the distribution (below a few micrometers; Supplementary Fig. S1), complete evaporation is expected to contribute increasingly to droplet removal, whereas optical shattering remains an efficient mechanism for larger droplets (Goffin et al, 2022).

The existence of this steady-state transmission plateau demonstrates that atmospheric laser clearing is not a transient phenomenon but can be continuously sustained under high-repetition-rate operation.

## Restoring optical transmission through dense fog

The steady-state transparent channel strongly enhances the transmission of the superimposed communication beam (Fig. 3). For initial fog attenuations ranging from 4 to 15 dB, the He-Ne transmission increases systematically with the clearing-laser pulse energy. For the densest fog investigated, corresponding to an initial attenuation of approximately 15 dB (3% linear transmission), the transmission enhancement reaches up to 12 dB at 130 mJ. Efficient clearing is therefore maintained even for randomly distributed multifilament bundles extending over several tens of meters.

The dependence of the transmission enhancement on the initial fog density provides further insight into the clearing efficiency (Supplementary Fig. S4). The enhancement varies almost linearly with the initial attenuation, with slopes of −0.72 at 80 mJ and −0.88 at 130 mJ. This behavior indicates that the clearing process remains effective without evidence of saturation over the investigated range of fog densities.

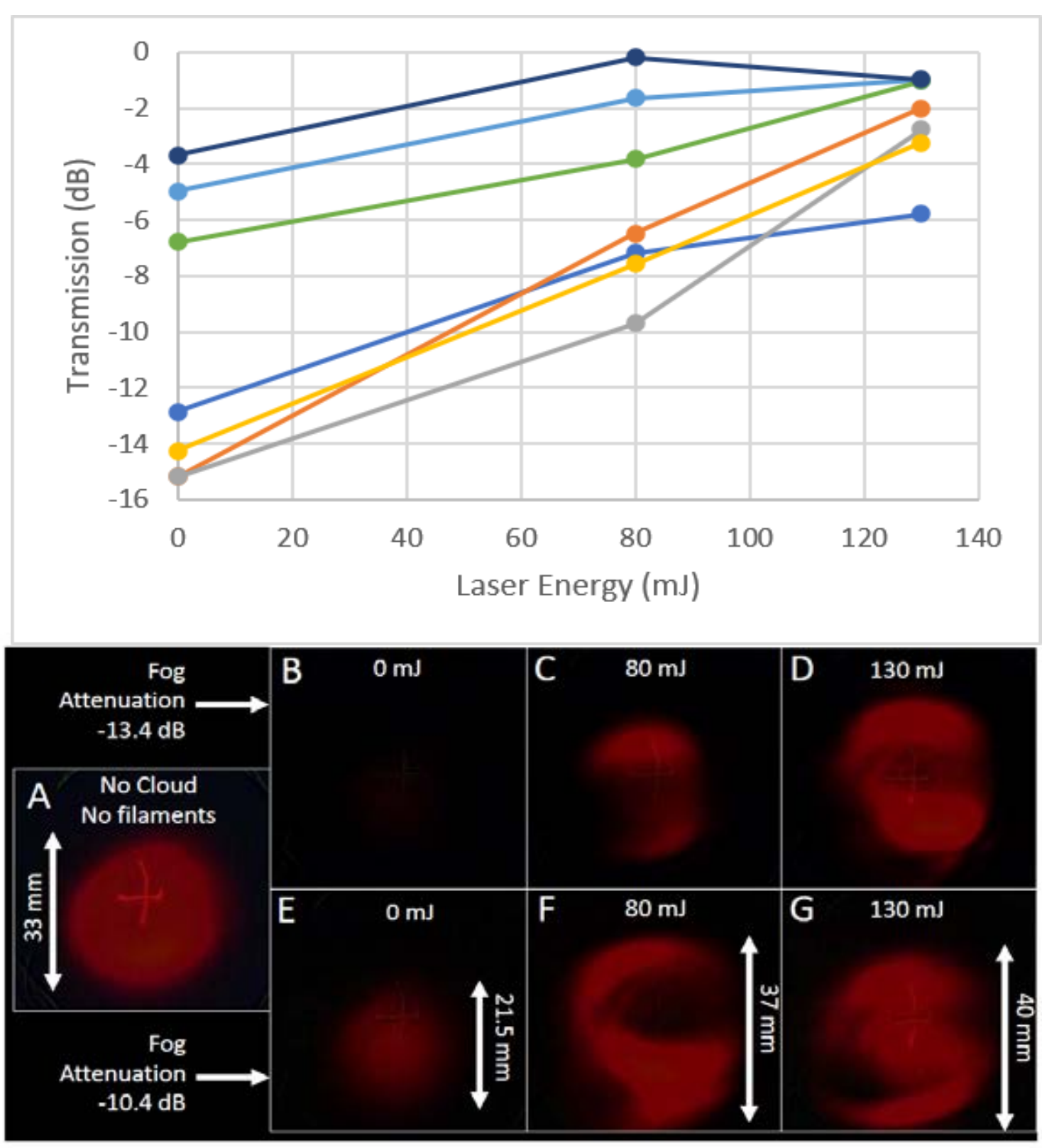

**Figure 3**: Upper : Transmission of the FSOC beam as a function of the energy of the clearing laser for different initial droplet concentrations, leading to attenuation that ranges from 4 dB to 15 dB (values at 0 mJ laser energy). Lower: images of the He-Ne beam at 140 m transmitted through 2 different fog densities and for different high-power laser energies. A video of the process is shown in supplementary S7.

Unlike the transmission of the clearing laser itself (Fig. 1), however, a clear difference remains between 80 and 130 mJ, particularly for the highest fog densities. This difference results from the finite 2-cm diameter of the communication beam. Whereas the high-power laser naturally clears the volume occupied by its own filament bundle, transmission of the He-Ne requires a sufficiently large fraction of its fixed cross-section to remain free of droplets. Increasing the clearing-laser energy therefore increases the fraction of the communication-beam cross-section that is effectively cleared, resulting in a higher overall transmission.

This substantial recovery of optical transmission is achieved with a very low distributed energy deposition along the propagation path. As shown in Supplementary Fig. S3, the total deposited energy increases almost linearly with the incident pulse energy, with measured slopes of 27% and 20% for the 100-m and 50-m focusing configurations, respectively. When normalized by the corresponding measured multifilament-bundle lengths of approximately 40 m and 15 m, these measurements yield longitudinally averaged deposited energies per unit length of 1.75 and 3.4 mJ $m^{-1}$, respectively. For a given focusing geometry, increasing the pulse energy primarily extends the multifilament region while leaving this longitudinally averaged deposition per unit length nearly unchanged. For the 100-m configuration, the bundle extends over 14 m at 80 mJ and 21.5 m at 130 mJ, while 25 and 38 mJ are deposited in air, corresponding to longitudinal averages of 1.78 and 1.81 mJ $m^{-1}$, respectively. These values are close to the 1.75 mJ $m^{-1}$ obtained at 240 mJ. Thus, over a threefold increase in pulse energy, the average energy deposited per unit length over the measured multifilament region remains essentially unchanged, while the length of this region increases substantially. These values are longitudinal averages obtained from the total deposited energy and the measured bundle length; the experiment does not provide a position-resolved measurement of the energy deposition along the bundle.

Despite this low longitudinally averaged energy deposition per unit length, operation at 1 kHz leads to a non-negligible cumulative thermal load. Inside the 3 m and 9 m fog chambers the average deposited power is approximately 6 W and 18 W.

The images in Fig. 3 provide a first indication of the resulting thermal effects on the communication beam. Although the transmitted He-Ne power is efficiently recovered, its spatial profile becomes increasingly distorted at high clearing-laser energies (see video Supplementary S7). At 130 mJ, the maximum beam diameter remains below approximately 4 cm, corresponding to an increase in divergence from 0.25 to 0.5 mrad. All transmission measurements were carried out by collecting the whole He-Ne beam, irrespective of its size. If the aperture of the detection optics were smaller than the beam diameter, cropping would introduce additional losses; since the transmitted beam diameter does not exceed approximately 4 cm even at the highest clearing-laser energy, the reported transmission values are unaffected by the observed beam broadening. No additional diffraction attributable to residual fog is observed, indicating that this beam spreading originates predominantly from laser-induced thermal blooming and refractive-index fluctuations. These thermal effects become increasingly important at longer interaction times; the additional influence of the confined experimental geometry is investigated below.

## Cloud clearing dynamics and thermal turbulence

On longer timescales, the clearing process exhibits a second, slower dynamics (Fig. 4a). Following the rapid establishment of the transparent channel, the transmission gain reaches a maximum of approximately 9 dB after 4 s, before progressively decreasing toward a quasi-steady-state value around 6 dB, while the clearing laser remains continuously on. A substantial transmission enhancement is maintained throughout this regime, showing that the transparent channel does not collapse, but that its long-time clearing efficiency is reduced.

This evolution occurs together with the thermal effects already visible in the He-Ne beam profiles of Fig. 3. Despite the low longitudinally averaged energy deposition per unit length, operation at 1 kHz results in a cumulative thermal load sufficient to generate refractive-index gradients, thermal blooming and beam wandering (Cheng et al., 2013; Schubert et al., 2016; Walch et al., 2021; Walch et al., 2024).

To identify the origin of the slow decline, we simulated the hydrodynamic response of the gas to this cumulative heating using a lattice Boltzmann method (Methods). The calculation contains no droplets and no optical propagation; it resolves the motion of the air alone. The heated core covers five numerical mesh cells in cross-sectional area, corresponding to an equivalent circular radius of 2.1 mm. This sits between the 1.2 mm diameter of ten close-packed 300 µm filaments and the bundle envelope measured on photographic paper, whose radius ranges from about 1.2 mm to 5 mm over the pulse energies (Fig. S2). The flow is confined by a cylindrical box with a 40 cm diameter. The results show that repeated heating of the millimeter-scale core on the axis drives a persistent buoyant plume, which the 40 cm chamber wall reorganizes into shear layers and recirculating cells extending over 10-15 cm, a large fraction of the tube cross-section (Fig. 4b). Another tube cross-section shortly after the laser onset is shown in Figure S5, and a plot of the velocity vectors corresponding to the velocity slice of Fig. 4b is provided in Figure S6. Time-animated sequences of the simulated laser induced turbulence are provided in Videos S8 (vorticity plot) and S9 (temperature plot).

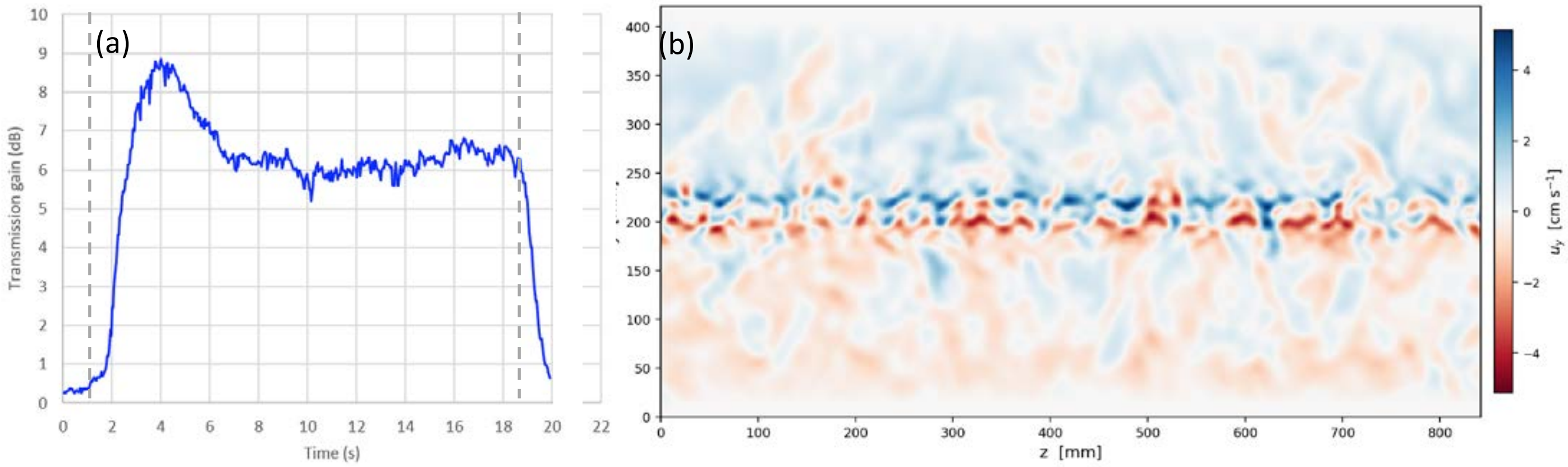

**Figure 4** (a) Experimental evolution of the transmission gain of the He-Ne communication beam. Vertical dashed lines indicate the switching on and off of the clearing laser. (b) Lattice Boltzmann simulation of the vertical gas velocity in a longitudinal plane of the 40-cm-diameter fog chamber at t = 6.5 s after the onset of the laser pulse train. Repeated laser heating generates a buoyancy-driven flow that is strongly modified by interaction with the chamber walls, leading to large-scale recirculation around the laser path.

A qualitative match between the simulation and the experimental data is observed, as the computed vertical velocities reach a few centimeters per second and the field is strongly asymmetric between the upper and lower halves of the tube, as buoyancy requires and as a cylindrically symmetric plume would not be. The response is also correctly ordered in time relative to the measurement: the clearing itself is complete within 1.5-2.4 s depending on pulse energy, whereas the hydrodynamic response needs several seconds to organize, comparable to the 4-8 s (Fig. 4a) over which the measured gain turns over toward its plateau. Once organized, the simulated flow appears statistically persistent over the simulated time window, with large-scale recirculating structures maintained around the laser path. This behavior is consistent with the experimentally observed degradation toward a lower transmission plateau, without collapse of the cleared channel.

The simulations therefore suggest that the long-time decrease in clearing efficiency observed in Fig. 4a is influenced by the confined experimental geometry. Wall-modified recirculation can transport droplets back toward the cleared region at the velocity required by the switch-off measurement, providing a plausible mechanism for the establishment of the lower steady-state transmission. Droplet transport is not explicitly included in the simulations, and this interpretation therefore follows from the calculated airflow rather than from a direct simulation of droplet dynamics.

**Discussion**

A key outcome of the present study is that efficient fog clearing can be sustained with a low longitudinally averaged energy deposition of only a few millijoules per meter over the multifilament region. For a fixed focusing geometry, increasing the pulse energy primarily extends the length of the multifilament bundle while leaving this longitudinal average nearly unchanged. For the 100-m focusing configuration, values of 1.78, 1.81 and 1.75 mJ m$^{-1}$ were obtained at 80, 130 and 240 mJ, respectively.

The nearly constant deposition per unit length is relevant when considering propagation over longer distances. Rather than increasing the longitudinally averaged deposited energy per unit length, additional pulse energy is observed to sustain filamentation over a longer propagation range (Rodriguez et al., 2004; Durand et al., 2013; Walch et al., 2023). If this behavior persists at higher pulse energies and appropriately scaled focusing conditions, distributing 0.5-1 J pulses over several hundred meters could provide transparent channels over distances relevant to atmospheric optical links. Achieving such scaling will require control of the longitudinal and transverse distribution of the multifilament bundle, for example through beam shaping, diffractive optics or wavefront engineering. The present measurements do not establish such scaling directly, but identify the low average energy deposition required per unit propagation length as a favorable starting point.

An indication that filamentation can persist through much more extended natural fog layers was obtained during the laser-guided lightning experiment at Säntis (Produit et al., 2021; Houard et al., 2023). During one lightning event under clear atmospheric conditions, laser guiding could be directly observed by optical cameras. During two additional events, dense fog prevented optical observation of both the laser beam and the lightning channel, while radio-frequency interferometric measurements nevertheless demonstrated laser guiding over approximately 60 m. Although no direct measurement of fog clearing was available in that vertical geometry, the observation demonstrates that the multifilament bundle remained capable of sustaining the interaction required for lightning guiding in dense fog over several tens of meters. In light of the present results, this is consistent with the progressive formation of a transparent propagation channel, although it does not constitute a direct measurement of clearing.

The experiments also identify limitations relevant to atmospheric implementation. First, the transparent channel is not established instantaneously. Under the dense-fog conditions investigated here, its formation requires typically several seconds, depending on pulse energy and fog density. This may constrain applications involving rapid beam repositioning, for example, when tracking aircraft or low-Earth-orbit satellites. Higher filament densities, repetition rates or optimized beam geometries may reduce this establishment time.

Second, cumulative energy deposition at kilohertz repetition rates inevitably heats the surrounding gas and produces refractive-index perturbations, resulting in thermal blooming and beam wandering of the communication beam (Cheng et al., 2013; Schubert et al., 2016; Walch et al., 2021; Walch et al., 2024). Unlike the recirculation observed in the chamber, these thermal effects are not specific to confinement and will remain relevant in open atmosphere. Adaptive wavefront correction may therefore be required for long-distance communication links. In the present experiment, the chamber walls additionally modify the laser-induced airflow, producing large-scale recirculation that can transport droplets back toward the cleared region and is consistent with the reduced long-time clearing efficiency observed in Fig. 4.

Atmospheric air motion introduces a separate constraint. For a cleared channel approximately 2 cm in diameter, a transverse wind of 20 m $s^{-1}$ displaces the surrounding air by 2 cm during the 1-ms interval between successive pulses, potentially replenishing the channel with droplets between laser shots. The relevant limit will depend on channel diameter, repetition rate, and droplet transport, and will be particularly important in strongly convective clouds. Conversely, increasing the repetition rate or the transverse extent of the filament bundle provides possible routes toward greater robustness against crosswinds.

In conclusion, we have demonstrated the formation and sustained maintenance of centimeter-scale transparent channels through dense fog using kilohertz multi-filamentation under atmospheric-scale propagation conditions. The resulting channels recover up to 12 dB of transmission for an independent optical communication beam. This performance is obtained with a longitudinally averaged energy deposition of only a few millijoules per meter over the multifilament region. For fixed focusing conditions, this average remains nearly unchanged as the pulse energy is increased, while the multifilament region extends over a longer distance. Increasing pulse energy therefore extends the cleared propagation range without a corresponding increase in the longitudinally averaged deposition per unit length. Remaining challenges include the initial channel-establishment time, thermal wavefront distortions, atmospheric crosswinds and the control of long multifilament bundles. High-repetition-rate filamentation can therefore sustain transparent channels through dense fog over extended propagation distances with low average energy deposition.

## Methods

### Experimental setup

The experiment was conducted in the 140 m-long hall of the former LAL (Laboratoire de l'Accélérateur Linéaire) facility in Orsay, France. The laser system was already described in Herkommer *et al.* (2020). It comprises a commercial 1030 nm seed oscillator, a fiber-Bragg-grating stretcher, a regenerative amplifier delivering 240 mJ pulses, and a multi-pass amplifier incorporating four industrial thin-disk gain modules that increase the pulse energy to approximately 800 mJ. The amplified pulses are subsequently recompressed from about 1 ns to the sub-picosecond regime using a folded Treacy compressor operated in a near-Littrow configuration. The system delivers up to 720 mJ per pulse at a repetition rate of 1 kHz, with a pulse duration of 920 fs and a beam quality factor $M^2$ = 1.4 (Herkommer et al., 2020).

This laser architecture combines near-terawatt peak power with kilowatt-class average power (Saraceno et al., 2019). The output beam was expanded and gently focused using a 40 cm-diameter, off-axis 8 × beam expander (Optical Surfaces), enabling precise positioning of the filamentation region and control over the diameter of the resulting filament bundle at any selected location along the propagation path. Part of the experimental campaign and of the data reported here was previously described in the doctoral thesis of one of the authors (Moreno, 2024).

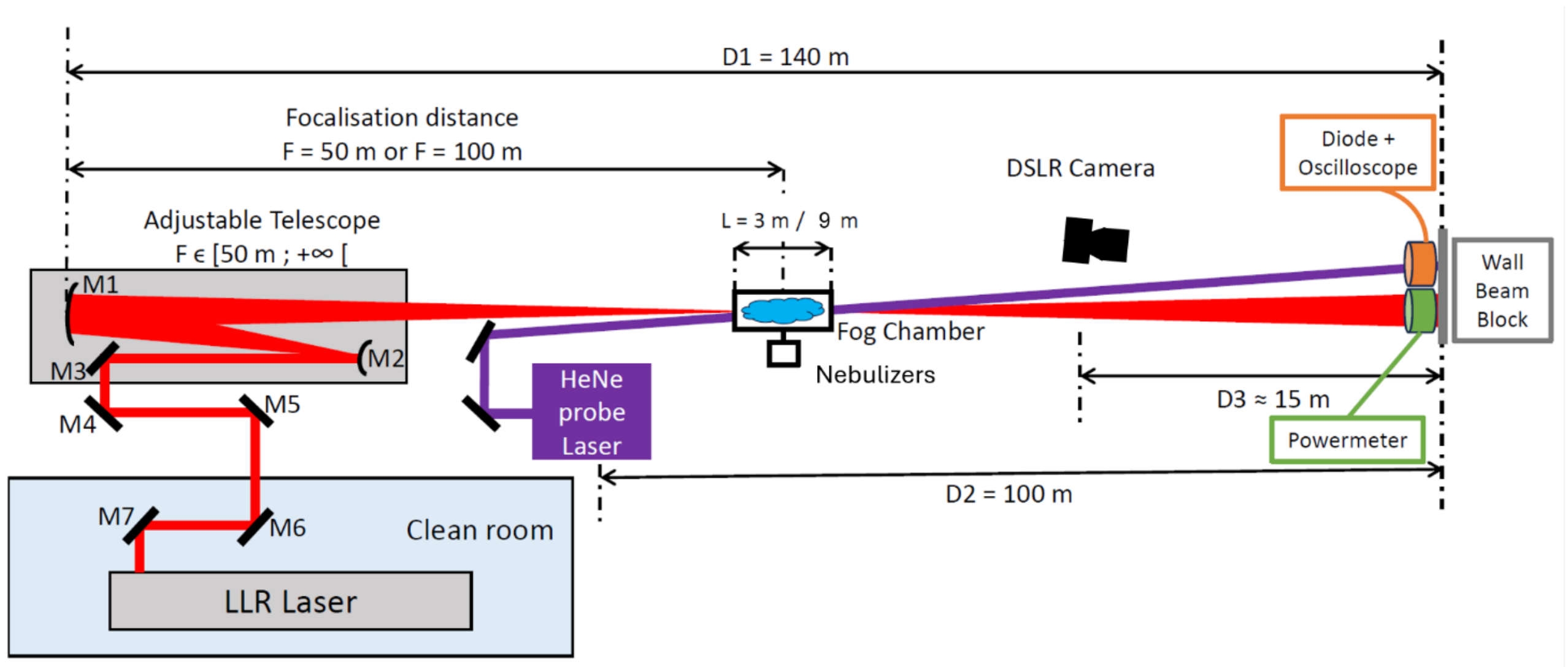

**Figure 5**: Experimental setup for long-range cloud-clearing experiments. The red beam represents the filamenting beam at 1030 nm. The filamenting beam is focused at 50 m or 100 m using a telescope. A semi-open cloud chamber of adjustable length from 3 to 9 m (Figure S1), is filled with fog by a set of atomizers, and moved along the optical path. A He-Ne beam is overlapped to the filamenting beam, serving as probe for FSO communications. Adapted from Moreno (2024).

Fog was generated using three ultrasonic nebulizers producing water droplets with a size distribution centered at 5-6 µm and a full width at half maximum (FWHM) of 3.5 µm, as characterized with a Grimm 1.109 aerosol spectrometer. To minimize the effects of air currents and turbulence, the fog-filled region was enclosed within tubes of 40 cm diameter, with adjustable lengths from 3 to 9 m. At each end, a cover plate with a 6 cm diameter eccentric hole (figure S1) was adjusted to restrict the diffusion out of the chamber. Buffer reservoirs were installed between the nebulizers and the tubes to homogenize the droplet concentration and improve long-term stability, typically over timescales of approximately 10 min. The fog density was continuously monitored through transmission measurements using a diode laser and photodiode positioned across the cloud volume.

An additional continuous-wave He-Ne laser operating at 632.8 nm (5 mW, Melles Griot), serving as a surrogate free-space optical communication beam, was co-propagated with the high-intensity Yb laser under a small crossing angle of ~1.5 mrad. After propagation through the fog, the transmitted He-Ne beam was characterized in terms of optical power, spatial profile, and pointing stability using a photodiode (with a spectral filter at 633 nm - 5 nm FWHM, and 2 removable neutral density filters OD1) and a DSLR camera (Nikon D810 equipped with a Zeiss 150 mm f/4 lens). The power of the high-average-power Yb laser was monitored along the propagation path using an air-cooled bolometric power meter (Ophir Starlite FL1100A-BB-65).

Assessment of the presence of filamentation, the surface of the filamentary region as compared to the whole beam size and the number of filaments contained in the bundle was achieved using impacts on the back side of photographic paper (Ilford Multigrade IV RC-44M) and subsequent analysis of the pictures. Whereas individual filaments can be clearly distinguished in the collimated regime, when focused they converge into a dense bundle, which in most cases prevents them from being distinguished on paper. When distinguishable, bundles typically contain 5-15 filaments of ~300 µm diameter, for energies ranging from 80 mJ (87 GW) to 240 mJ (260 GW). As the critical power $P_c \approx 4.9$ GW under our conditions (Walch et al., 2023), approximately 3.5 $P_c$ are required to sustain the propagation of a single filament, which is typical for this loosely focused regime. The transverse dimensions of the whole beam and of the filament bundles at the location of the fog chambers are shown in Supplementary Fig. S2.

**Simulations of the establishment dynamics of the transparent channel**

Numerical simulations were performed to model the initial clearing phase, during which direct laser-droplet interactions, dominated by plasma-mediated shattering for sufficiently large droplets, drive the initial clearing. Energy losses in the laser-droplet interaction are assumed to arise primarily from plasma formation (75-80%), with the remaining fraction attributed to Mie scattering and linear absorption (Jeon et al., 2015; Wolf, 2018). Plasma formation within droplets and the subsequent explosive shattering induced by laser filaments have been extensively investigated (Courvoisier et al., 2003a; Favre et al., 2002; Lindinger et al., 2004; Peña et al., 2021; Goffin et al., 2022), with typical breakdown thresholds on the order of $10^{12}$ W cm$^{-2}$. Reported thresholds lie in the range $3 \times 10^{11}$-$10^{12}$ W cm$^{-2}$ for 100 fs pulses; for the longer pulses used here a lower threshold is expected, since laser-induced breakdown in water shows a factor of about two between 100 fs and 1 ps (Linz et al., 2025), and a value of $10^{12}$ W cm$^{-2}$ was therefore adopted. As the intensity of the photon bath surrounding the filaments is about one order of magnitude below this threshold (1-2 × $10^{11}$ W cm$^{-2}$), whereas the intensity within the filaments is clamped at approximately $5 \times 10^{13}$ W cm$^{-2}$, shattering is dominated by the interaction with the filaments themselves. Based on the analysis of laser-induced burns on photographic paper, the filament bundle was assumed to consist of ten filaments for the 130 mJ case and five filaments for the 80 mJ case, each with a diameter of 300 µm. For each laser pulse, the number of droplets intercepted by the filament bundle was calculated, together with the corresponding reduction in droplet number density. As the laser intensity decreases along the propagation path because of interaction with the cloud, droplets were assumed to undergo plasma-mediated shattering only while the local filament intensity remained above $10^{12}$ W cm$^{-2}$.

The droplet number density therefore progressively decreases with successive laser pulses, leading to a corresponding increase in the transmission of the He-Ne probe beam. During this initial clearing phase, droplet replenishment within the cleared channel was neglected. Despite its simplicity, the model reproduces the experimentally observed early-time dynamics with good agreement.

**Lattice Boltzmann Simulations**

The long-time reduction of the transmission gain suggests a mechanism distinct from droplet removal. We therefore simulated the gas-dynamic response to the cumulative heat deposited by the pulse train, in order to test whether kilohertz heating alone can generate chamber-scale motion on the observed timescale. The simulations model neither filament propagation, nor droplet shattering, nor optical transmission.

The flow is computed with a double-distribution lattice Boltzmann method (LBM) on a D3Q39 lattice, in which a primary population carries mass, momentum and translational energy while a secondary population carries the remaining internal degrees of freedom, so that air is represented as a polyatomic gas with $\gamma = 1.4$. The equilibrium populations are obtained from a 13-moment numerical equilibrium rather than a low-order polynomial expansion, which improves robustness for compressible thermal flows. The macroscopic equations recovered by the LBM are described in the general framework of Krüger et al. (2017), while the recovery of the Navier-Stokes-Fourier equations with the present numerical-equilibrium formulation and its GPU implementation are detailed in Latt et al. (2020) and Coreixas and Latt (2025). Collisions use a single relaxation time $\tau = \nu/T + \Delta t/2$ (Bhatnagar et al., 1954; Krüger et al., 2017), where the kinematic viscosity $\nu$, the temperature T, and the timestep $\Delta t$ are expressed in physical units. Buoyancy is applied as a vertical momentum source proportional to the local density deficit,

$$M_y \leftarrow M_y + \tau(\rho_0 - \rho)g,$$

where $\rho_0 = 1.21\ \mathrm{kg/m^3}$ at ambient conditions, $\rho$ is the local density, and $g = 9.81\ \mathrm{m\ s^{-2}}$ is the magnitude of gravitational acceleration which points downward in the y-direction.

The computational domain is a longitudinal segment of the fog chamber, 0.42 × 0.42 × 0.84 m, discretized on 256 × 256 × 512 cells at a uniform spacing $\Delta x = 1.64$ mm with a time step $\Delta t = 4.0$ µs. The 40 cm chamber wall is resolved by 243 cells and imposed as a no-slip boundary by halfway bounce-back of the incoming populations. The domain is periodic along the tube axis, so the model represents a longitudinal section rather than the full 3 m chamber and has no end walls. A selective spatial filter provides numerical stability. The simulation is initialized with quiescent air at $T_0 = 300$ K and advanced to 13.3 s of physical time.

The laser is represented as a heated cylindrical core on the axis, of effective radius $r_{eff} = 2.1$ mm and cross-sectional area $A = 1.35 \times 10^{-5}\ \mathrm{m^2}$. Within it the target temperature is raised from $T_0$ toward $T_{max} = 450$ K through a Gaussian temporal envelope, the imposed density being adjusted at constant pressure to avoid an artificial pressure discontinuity. This yields a period-averaged overheat $\langle \Delta T \rangle \simeq 3.8$ K and a deposited power of 3.4 W $\mathrm{m^{-1}}$. A weak helical modulation of the heated core, of amplitude below 1 % and of axial wavelength comparable to the domain length, breaks the translational and mirror symmetry of the configuration.

**Use of generative AI**

ChatGPT (OpenAI) was used to assist with language editing and stylistic refinement of the manuscript. All scientific content, interpretations and conclusions were provided, reviewed and approved by the authors.

## References

Ali Reza SB et al., Generation of multiple obstruction-free channels for free space optical communication, Optics Express **31**, 3168-3178 (2023)

Bergé L et al., Ultrashort filaments of light in weakly-ionized, optically-transparent media, *Rep. Prog. Phys.* **70**, 1633-1713 (2007)

Bhatnagar PL, Gross EP, Krook M, A model for collision processes in gases. I. Small amplitude processes in charged and neutral one-component systems, *Phys. Rev.* **94**, 511-525 (1954)

Chen YA. et al., An integrated space-to-ground quantum communication network over 4,600 kilometres. *Nature* **589**, 214–219 (2021).

Cheng YH, J. K. Wahlstrand, N. Jhajj, H. M. Milchberg, The effect of long timescale gas dynamics on femtosecond filamentation, *Opt. Express* **21**, 4740 (2013)

Chin SL et al., The propagation of powerful femtosecond laser pulses in optical media: physics, applications, and new challenges, *Can. J. Phys*. **83(9),** 863-905 (2005)

Coreixas C, Latt J, GPU-based compressible lattice Boltzmann simulations on non-uniform grids using standard C++ parallelism: From best practices to aerodynamics, aeroacoustics and supersonic flow simulations, *Comput. Phys. Commun.* **317**, 109833 (2025)

Couairon A, A Mysyrowicz, Femtosecond filamentation in transparent media, *Phys. Rep.* **441**, 47-189 (2007)

Courvoisier F et al., Ultraintense Light Filaments Transmitted through Clouds, *Appl.Phys.Lett.* **83**, 213-215 (2003b)

Courvoisier F et al., Plasma Formation Dynamics within a Water Microdroplet at Femtosecond Timescales *Opt.Lett.* **28(3)**, 206-208 (2003a)

De La Cruz L et al., High repetition rate ultrashort laser cuts a path through fog. *Applied physics letters* **109**,251105 (2016).

Durand M et al., Kilometer range filamentation, *Optics Express* **21**, 26836-26845 (2013)

Frigerio et al., Filament-induced breakdown spectroscopy of solids through highly scattering media, *Optics Letters* **49**, 4942-4945 (2024)

Favre C et al., White-light Nanosource with Directional Emission, *Phys.Rev.Lett.* **89(3)** 035002 (2002)

García-García F et al., Fine-scale measurements of fog-droplet concentrations: a preliminary assessment, *Atmospheric Research* **64**, 179–189 (2002)

Goffin et al., Atmospheric Aerosol Clearing by Femtosecond Filaments, *Phys. Rev. Applied* **18**, 014017 (2022)

Herkommer C *et al.*, Ultrafast thin-disk multipass amplifier with 720 mJ operating at kilohertz repetition rate for applications in atmospheric research, *Optics Express* **28**, 30164 (2020).

Houard A *et al.*, Study of filamentation with a high power high repetition rate ps laser at 1.03 µm, *Optics Express* **24**, 7437-7448 (2016)

Houard A *et al.*, Laser-guided lightning, *Nature Photonics* **17**, 231 (2023).

Jahid A *et al.*, A contemporary survey on free space optical communication: Potentials, technical challenges, recent advances and research direction, *J. of Network and Computer Applications* **200**, 103311 (2022)

Jeon C *et al.*, Interaction of a single laser filament with a single water droplet, *J. Opt* **17**, 055502 (2015)

Karlsson KG, Devasthale A, Inter-Comparison and Evaluation of the Four Longest Satellite-Derived Cloud Climate Data Records: CLARA-A2, ESA Cloud CCI V3, ISCCP-HGM, and PATMOS-x, *Remote Sens.* ***10*****(10)**, 1567 (2018)

Kasparian J *et al.*, White-light filaments for atmospheric analysis, *Science* **301** (5629), 61-64 (2003)

Krüger T et al., The Lattice Boltzmann Method: Principles and Practice, Graduate Texts in Physics, Springer (2017)

Lahav O *et al*., Long-lived waveguides and sound-wave generation by laser filamentation, *Phys.Rev.*A **90**, 021801 (2014).

Latt J *et al.*, Efficient supersonic flow simulations using lattice Boltzmann methods based on numerical equilibria, *Phil. Trans. R. Soc.* A **378**, 20190559 (2020).

Lindinger et al., Time-resolved explosion dynamics of H2O droplets induced by femtosecond laser pulses, *Applied Optics* **43**, 5263 (2004)

Linz et al., Laser-induced plasma formation and cavitation in water: from nanoeffects to extreme states of matter, *Rep. Prog. Phys.* **88**, 088501 (2025)

Loescher R et al., High-power sub-picosecond filamentation at 1.03 µm with high repetition rates between 10 and 100 kHz, *APL Photonics* **8**, 111303 (2023)

Méjean G et al., Multifilamentation Transmission through Fog, *Phys.Rev.* E. **72** 026611 (2005)

Moreno V, Laser-guided lightning: atmospheric discharges controlled by high-repetition rate laser filaments, doctoral thesis No. Sc. 5816, Université de Genève (2024), doi:10.13097/archive-ouverte/unige:180749

Muhammad S et al., Characterization of fog attenuation in terrestrial free space optical links, *Optical Engineering* **46**(6), 066001 (2007)

Peña et al., Shockwave enhancement from temporally separated filaments interacting with a water droplet, *Journal of the Optical Society of America* B **38**, 2437-2442 (2021)

Point G et al., Superfilamentation in Air*, Physical Review Letters* **112**, 223902 (2014).

Produit T *et al.*, The laser lightning rod project, *Eur. Phys. J. Appl. Phys*. **93**, 10504 (2021).

Rodriguez M et al., Kilometer-Range Non-linear Propagation of fs Laser, *Phys. Rev.* E **69** , 036607 (2004)

Saraceno C et al., The amazing progress of high-power ultrafast thin-disk lasers, *J. Eur. Opt. Soc.* **15**, 15 (2019).

Schimmel, G et al. Free space laser telecommunication through fog. *Optica* **5**, 1338–1341 (2018)

Schroeder MC et al., Molecular quantum wakes for clearing fog, *Optics Express* **28**, 11463-11471 (2020)

Schroeder MC et al., Opto-mechanical expulsion of individual micro-particles by laser-induced shockwave in air, *AIP Advances* **12**, 095119 (2022)

Schubert E et al., Dual-scale turbulence in filamenting laser beams at high average power, *Physical review* A **94**, 043808 (2016).

Walch P et al., Cumulative air density depletion during high repetition rate filamentation of femtosecond laser pulses: Application to electric discharge triggering, *Appl. Phys. Lett.* **119**, 264101 (2021)

Walch et al., Long distance laser filamentation using Yb:YAG kHz laser*, Sci Rep.* **13(1),** 18542 (2023)

Walch P et al., Impact of gravitational force on high repetition rate filamentation of femtosecond laser pulses in the atmosphere, *Appl. Phys. Lett.* **124**, 151101 (2024)

Wang et al., Structured light signal transmission through clouds, *J. Appl. Phys.* **133**, 043102 (2023)

Wolf, JP, Short-pulse lasers for weather control, *Rep. Prog. Phys*. **81,** 026001 (2018).

Yan B et al., Laser-filamentation-assisted 1.25 Gb/s video communication under harsh conditions, *Optics and Laser Technology* **131**, 106391 (2020)

Zhang C et al., Optical breakdown during femtosecond laser propagation in water cloud, *Optics Express* **27,** 8456-8475 (2019)

## Supplementary information

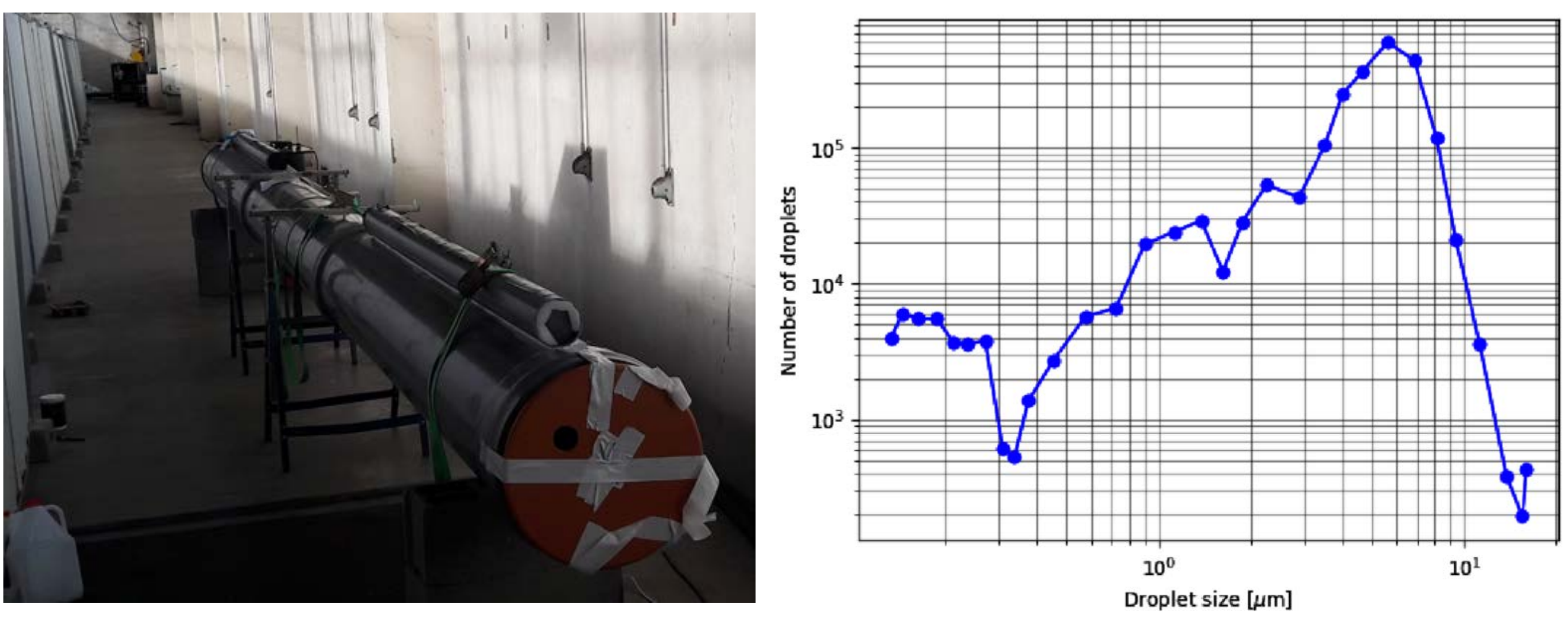

**Figure S1**: Semi-open cloud chamber of 9 m length; right: Droplet size distribution of the fog, measured by a Grimm 1.109, from (De la Cruz et al 2016)

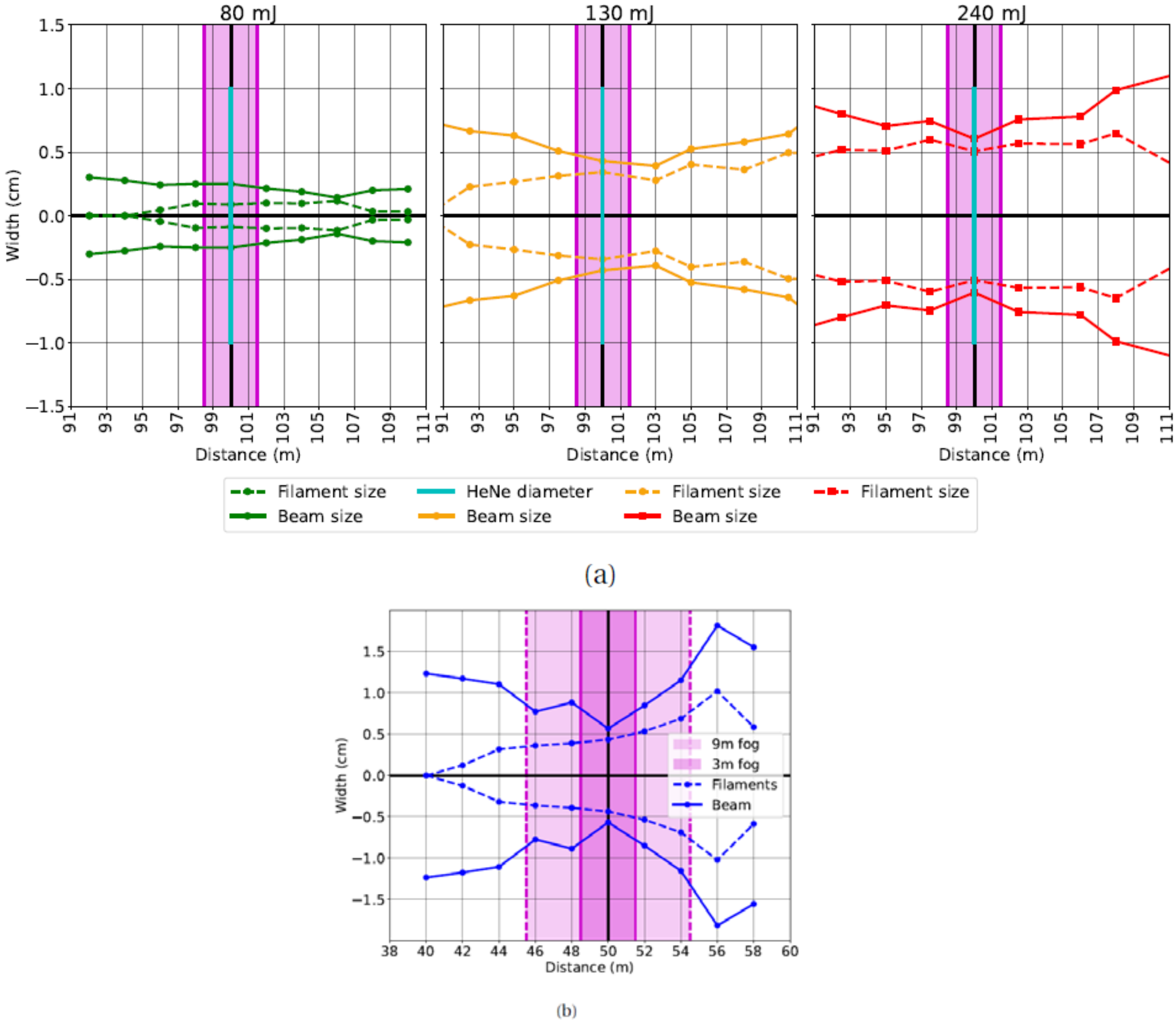

**Figure S2**. Dimensions of the whole beam and the filament bundles using impacts on photographic paper at the location of the fog chambers for the (a) 100 m focusing and (b) 50 m focusing cases

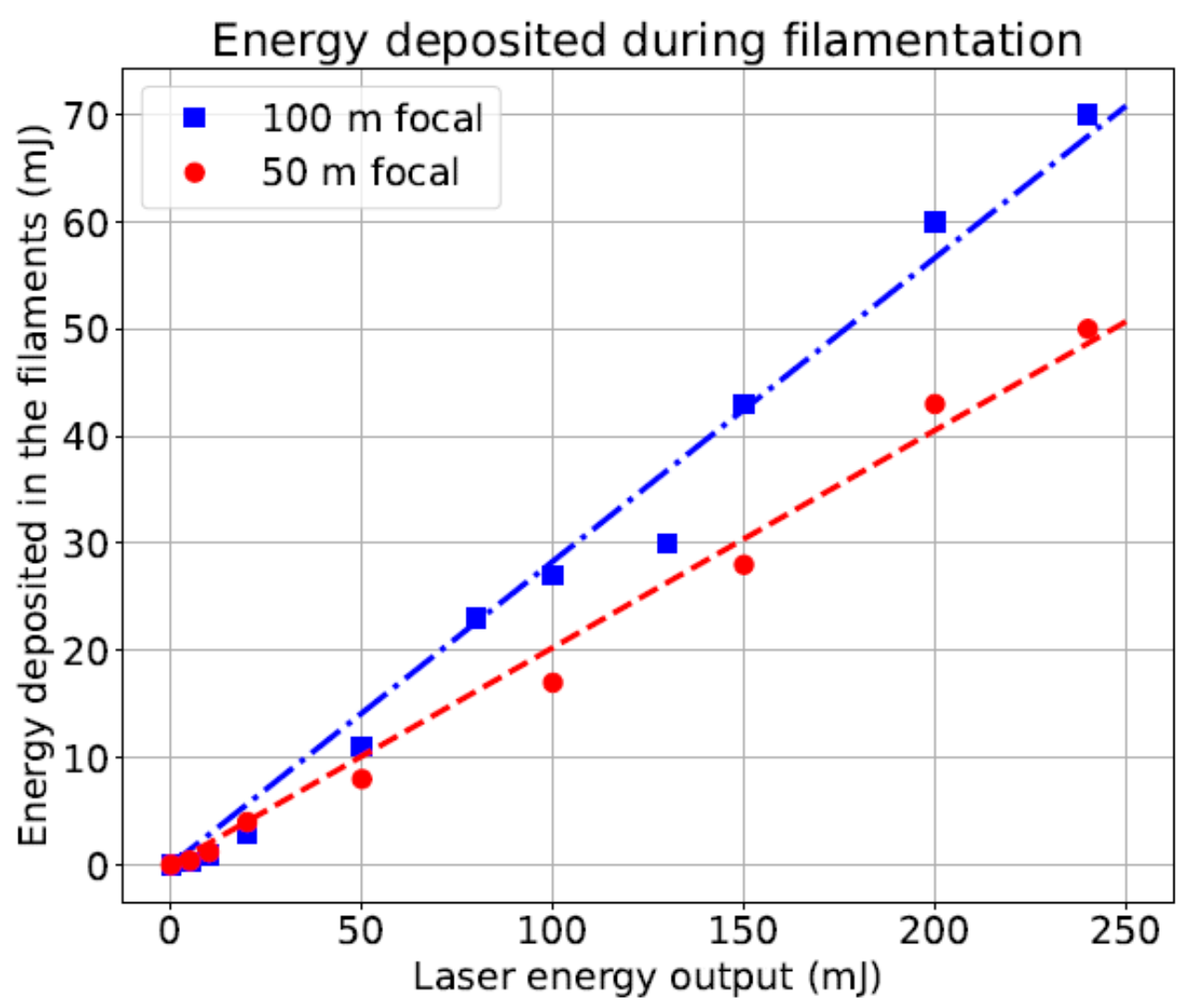

**Figure S3**. Energy deposition by the filament bundle in air as a function of initial energy and focusing distance. Lines are the linear fit across the data

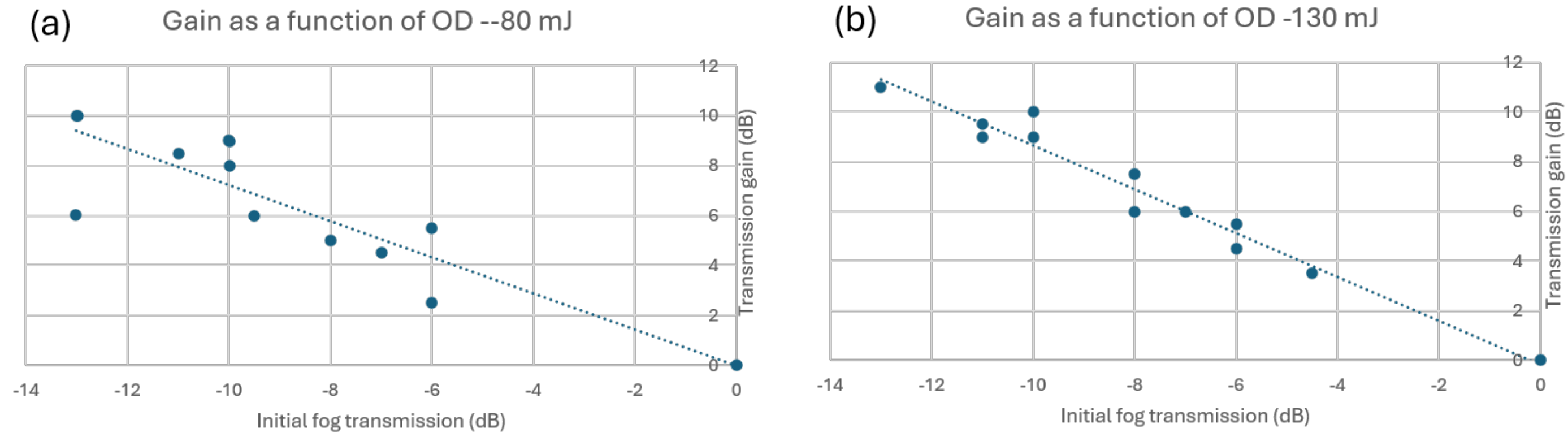

**Figure S4**: Transmission Gain for laser energies of (a) 80 mJ and (b) 130 mJ. Vertical scale is the difference in ($T_{dB}$(laser) - $T_{dB}$(no laser)) and horizontal scale is $T_{dB}$(no laser), with $T_{dB}$ the transmission in dB.

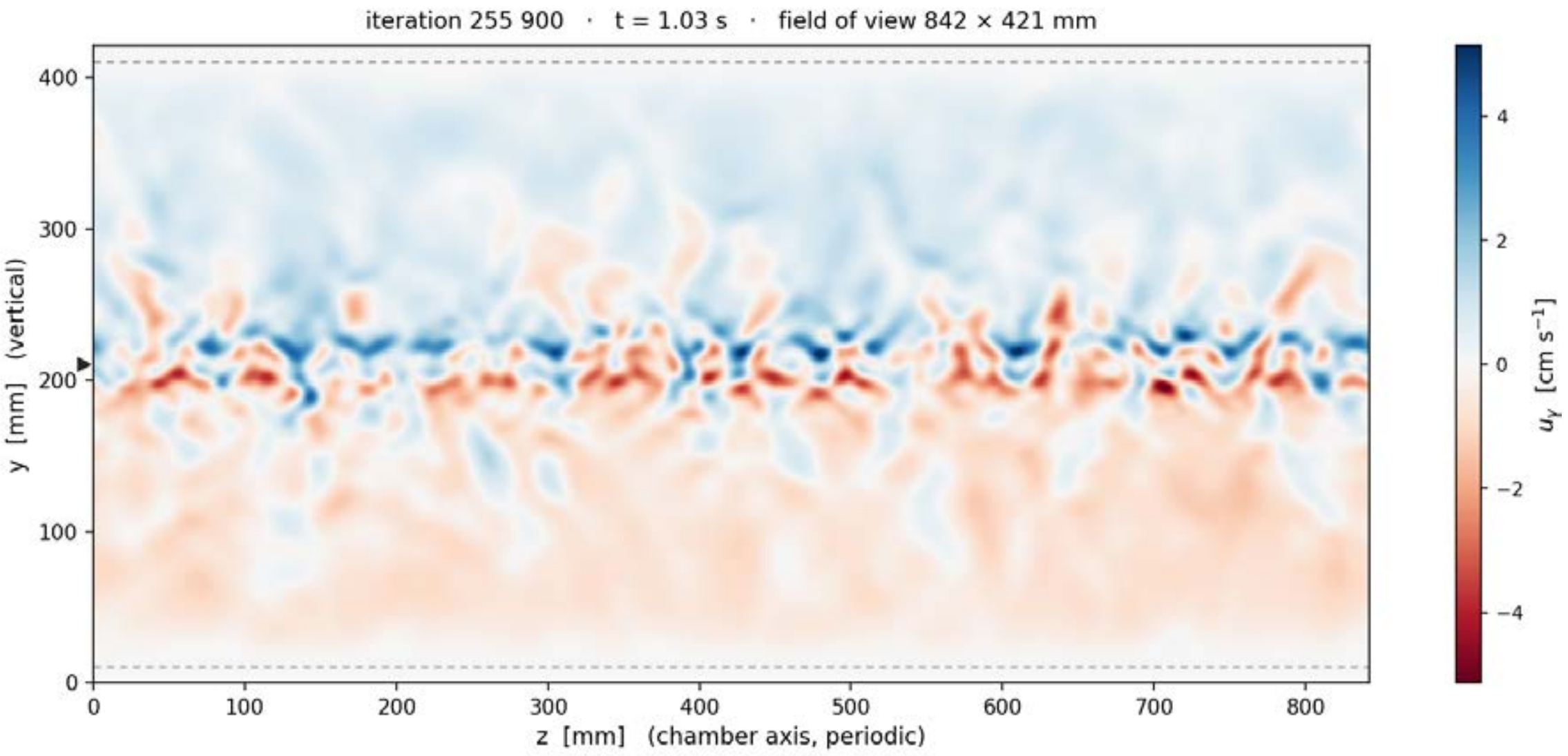

**Figure S5**: Vertical gas velocity $u_y$ at t = 1.03 s - the same quantity and colour scale as Fig. 4b, shortly after onset.

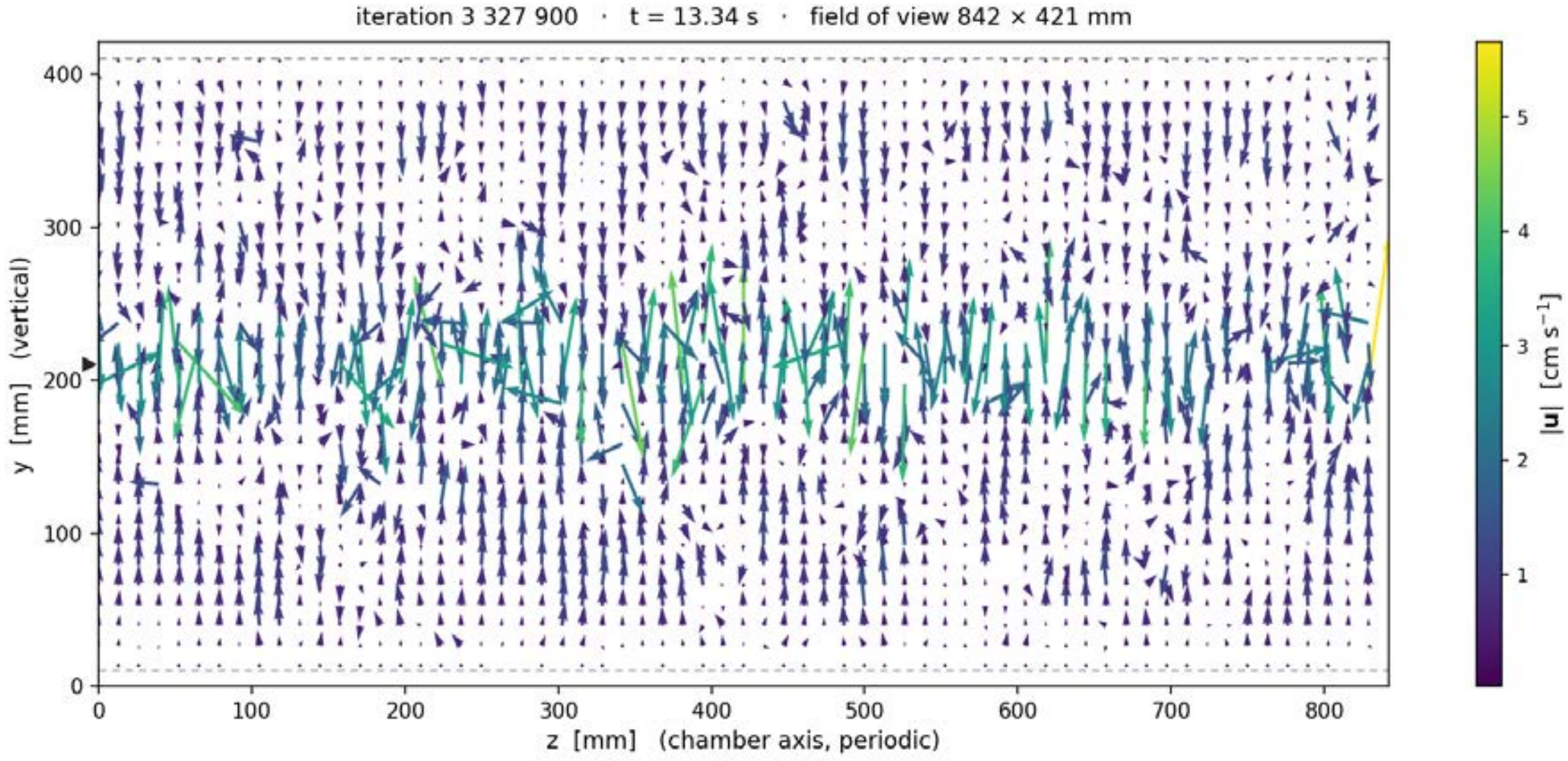

**Figure S6:** Velocity vectors at t = 13.3 s, the instant of Fig. 4b, showing the sense and closure of the motion whose vertical component Fig. 4b resolves.

**Video S7**: Measurement of the turbulence induced by the clearing laser on the transmitted signal.

**Video S8**: Lattice Boltzmann simulation of the air motion (vorticity plot) induced by the filament bundle in the cloud chamber, longitudinal and transverse cut.

**Video S9**: Lattice Boltzmann simulation of the air motion (temperature plot) induced by the filament bundle in the cloud chamber, frontal view.